# IoT-Flock: An Open-source Framework for IoT Traffic Generation

Syed Ghazanfar, Faisal Hussain
*Al-Khawarizmi Institute of Computer
Science (KICS) Lahore, Pakistan*
ghazanfar.abbas@kics.edu.pk,
faisal.hussain.engr@gmail.com

Atiq Ur Rehman, Ubaid U. Fayyaz
*Al-Khawarizmi Institute of Computer
Science (KICS) Lahore, Pakistan*
atiq.rehman@kics.edu.pk,
Ubaid@uet.edu.pk

Farrukh Shahzad, Ghalib A. Shah
*Al-Khawarizmi Institute of Computer
Science (KICS) Lahore, Pakistan*
farrukh.shahzad@kics.edu.pk,
ghalib@kics.edu.pk

*Abstract*—Network traffic generation is one of the primary techniques that is used to design and analyze the performance of network security systems. However, due to the diversity of IoT networks in terms of devices, applications and protocols, the traditional network traffic generator tools are unable to generate the IoT specific protocols traffic. Hence, the traditional traffic generator tools cannot be used for designing and testing the performance of IoT-specific security solutions. In order to design an IoT-based traffic generation framework, two main challenges include IoT device modelling and generating the IoT normal and attack traffic simultaneously. Therefore, in this work, we propose an open-source framework for IoT traffic generation which supports the two widely used IoT application layer protocols, i.e., MQTT and CoAP. The proposed framework allows a user to create an IoT use case, add customized IoT devices into it and generate normal and malicious IoT traffic over a real-time network. Furthermore, we set up a real-time IoT smart home use case to manifest the applicability of the proposed framework for developing the security solutions for IoT smart home by emulating the real world IoT devices. The experimental results demonstrate that the proposed framework can be effectively used to develop better security solutions for IoT networks without physically deploying the real-time use case.

*Index Terms*—Traffic Generator, IoT Traffic Generator, IoT Flock, IoT Use Case, Intrusion Detection System, IoT Security

## I. INTRODUCTION

Internet of things (IoT) has recently induced as a topic of intense interest among the research community since it integrates various technologies. The main concept of IoT is that various devices comprising different technologies will be connected and communicating with each other without human intervention. IoT is a communication paradigm that gives the concept of communication between the objects of our daily life, connected over the internet. IoT has gained the capability of interacting with a wide variety of devices such as household appliances, industrial machines, robots, drones, power generation systems and many others. By controlling and managing a massive amount of data, produced by such devices, IoT can provide new services to luxuriate human life.

In the current era, security is the major concern of IoT [1]. Firewalls, intrusion detection systems (IDS) and intrusion prevention systems (IPS) are the major security shields to protect the devices and network from cyber-attacks. Most of the firewalls, IDS and IPS filter the normal and malicious traffic based upon the signatures, i.e., static predefined rules. While a few IDS and IPS use artificial intelligence (AI) techniques along with signatures to detect the attack traffic. The IDS and IPS that filter the intrusive attempts using both signatures and AI techniques are more effective as compared to those which only use signatures. The AI-based IDS and IPS are trained and tested using normal and attack traffic datasets. These datasets are collected by two approaches, i.e., either by using real systems to generate malicious and normal network traffic or by using some traffic generator tools which mimic the real-time network traffic.

No matter, the present IDS and IPS technology is quite mature but it is inadequate for IoT Systems [2]. The primary cause is the communication protocols like CoAP, MQTT, etc., which IoT devices use, are not employed in a traditional network as different protocols carry different vulnerabilities and requirements [2]. Another crucial factor is the limited processing and storage capacity of the IoT devices due to which host-based IDS cannot be installed on IoT devices. However, the network-based IDS can protect the IoT network and devices from cyber-attacks if they are equipped with the support of IoT protocols.

There exist some datasets like KDD-99 [3], NSL-KDD [4], CAIDA [5], ISCX [6], etc., that are widely used for developing the security systems to protect the IoT networks from malicious attacks. However, these datasets have certain issues with respect to IoT, like these datasets don't have the traffic of commonly used IoT protocols, e.g., MQTT, CoAP, etc. Moreover, some of these datasets are so old that they are outdated, as there is a quite difference between the past and current cyber-attacks [7]. Nevertheless, this dilemma can be untangled by generating the dataset through a network traffic generator tool which can generate both normal and attack traffic of commonly used IoT protocols.

A network traffic generator tool is a kind of software that allows a user to generate the detailed custom packets. The traffic generator tools are extensively used by researchers and secu-rity providers in order to develop and test security applications like IDS, IPS, etc. Moreover, it can be used for the evaluation of network performance like stress testing [8]. Furthermore, network engineers use traffic generator tools for benchmarking the network features and to troubleshoot the network problems. So far, many traffic generator tools/frameworks have been proposed [9]–[13] by both the research and software development community. However, these frameworks/tools have certain shortcomings like IoT application layer protocols support is still missing in these tools. Similarly, most of the traffic generating tools lack of generating the attack traffic. Hence, the existing traffic generator tools/frameworks are inadequate for developing and testing the security solutions of the IoT networks. Therefore, we proposed a framework for IoT traffic generation which can generate both the normal and attack traffic of two widely used IoT application layer protocols, i.e., MQTT and CoAP.

The main focus of this work is to propose a framework which consists of an IoT traffic generator tool so that the researchers may easily build their own use case, model IoT devices into it and then generate & analyse the traffic of the use case in order to develop better security solutions for IoT. The proposed traffic generation framework can also be used in stress testing of different IoT-based network utilities like switches, routers, etc., by generating a large amount of IoT device traffic. Moreover, it can also be used for the designing and testing of IoT security providing entities like IDS, IPS, etc. The key contributions of this work are as follows:

- We proposed an open-source framework which consists of an IoT traffic generator tool which is capable of generating IoT normal and attack traffic over a real-time network using a single physical machine.
- To our best knowledge, we are the first to design an open-source IoT traffic generator which supports two application layer IoT protocols, i.e., MQTT and CoAP.
- We devised IoT device modelling by introducing the concept of time profile and data profile in order to better emulate the IoT devices.
- Furthermore, we implemented a real-time smart home use case using the proposed IoT traffic generation framework and demonstrated how the generated traffic can be used to develop a machine learning based security solution.

The rest of the paper is structured as follows: Section II presents a review of some existing traffic generator tools/frameworks. Section III describes the features, architecture and working of the proposed framework. Section IV discusses the experimentation and demonstrates how IoT traffic can be generated and used for developing machine learning based security solutions. Lastly, Section V concludes the paper.

## II. RELATED WORK

The network traffic generator tools are used extensively for evaluation of network performance like throughput calculation, stress testing [8], etc. So far, many traffic generator tools have been proposed in the literature. In [14], authors proposed a traffic generator tool for switch testing. The developed traffic generator consists of both hardware and software modules. The software module generates configurations and parameters according to the traffic model selected by the user. While the hardware module generates the packets as specified by the software module and send it to the network interface module. In [11], authors introduced a scriptable traffic generator which consists of both hardware and software modules. The software module was developed for packet configurations while the hardware module was developed to control the packets rate and latency. However, it can only generate ICMP, ARP, TCP, UDP and IP protocols traffic.

In [8], authors proposed a traffic generation framework for testing the deep packet inspection (DPI) tools. The proposed framework generates network traffic based on user behaviour emulation. They gathered the real-time traffic, analyzed it and extracted the typical user behaviour to emulate it later for testing the DPI tools. In [12], authors designed a hardware IoT device which can generate the IoT traffic. The designed hardware can generate traffic flows simultaneously based on the interval length and data size. However, this device only generates layer 2 traffic. While in case of IoT, we are primarily concerned with application layer protocols like MQTT, CoAP, HTTP etc. In [15], authors proposed a traffic generator framework by integrating the machine type communication (MTC) traffic models with big data. The framework was proposed to evaluate the performance of mobile networks.

In [13], authors designed a tool that not only generates the network traffic but also evaluates the network performance as well as functional testing at the switch level. The tool can be used to generate different test scenarios and analyze the response in order to check the functional aspects of the switch network. In [16], authors proposed a traffic generator which can generate the network traffic of the network layer and transport layer. The traffic generator supports TCP and UDP protocols. A user has to specify the content, data size, duration, flow features and traffic model in order to generate the traffic.

In addition to the literature, many open-source traffic generator tools are developed by the software development community. For example, Packet Sender [9] is an open-source tool that supports sending and receiving TCP, UDP and SSL traffic in order to test network APIs and network connectivity. Likewise, D-ITG tool [10] emulates the network traffic of TCP, UDP, ICMP protocols. Moreover, it also measures the network performance metrics like throughput, delay, losses, etc., based upon the network flows.

Although, many traffic generator tools/frameworks have been developed by both the research and software development community IoT application layer protocols support is still missing. Furthermore, a large number of network traffic generator tools do not generate the attack traffic. Hence, the existing traffic generator tools are inadequate for testing the performance and security of the IoT networks. Therefore, we proposed a novel open-source framework which consists of a traffic generator tool. The proposed framework can generate the two widely used IoT application layer protocols, i.e., MQTT and CoAP. Moreover, it can generate the normal and attack traffic over a real-time network which is imperative for testing and evaluating the security solutions proposed for IoT.

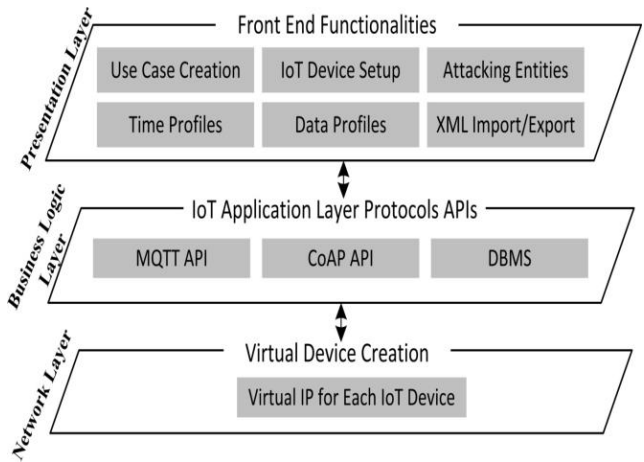

Fig. 1. Layer-wise Core Functionalities of IoT-Flock

### III. PROPOSED FRAMEWORK

The proposed framework consists of an open-source IoT traffic generator tool which can generate the normal and attack traffic of IoT devices over a real-time network. We named it as IoT-Flock. Fig. 1 shows layer-wise core functionalities of the IoT-Flock. The IoT-Flock has following distinct features as compared to other commercially or publicly available traffic generator tools:

- IoT-Flock has two working modes, i.e., GUI mode and console mode which allows a user to create real-time IoT use cases add thousands of IoT devices into the use case.
- IoT-Flock provides XML support to import or export the designed use case. Thus, a user can create, share and run a use case through an XML file.
- IoT-Flock can also generate MQTT and CoAP related attacks. To our best knowledge, this feature is not supported yet by any other open-source IoT traffic generator tool. Thus, a user can easily create both normal and attacking devices in the same use case and generate their traffic.

#### A. IoT-Flock Architecture

We followed layered architecture in order to develop the IoT-Flock. IoT-Flock comprises of three layers, i.e., presentation layer, business logic layer and network layer. The following subsections describe these layers.

*1) Presentation Layer:* The presentation layer is responsible for viewing the interface to the end-user. It works in two modes, i.e., console mode and graphical user interface (GUI) mode. In console mode, a user will import the use case XML file through the console and run the tool for generating the IoT traffic as defined in the XML file. In GUI mode, a user will first create an IoT use case then add different IoT devices into it. A user can add MQTT and CoAP devices into the use case. While creating a device, a user has to give the information about the device name, device IP address, device type, number of devices and protocol type in order to emulate the device traffic over the real-time network using IoT-Flock.

If a user selects MQTT protocol, then the MQTT GUI form will be enabled and the user has to give the further device information of MQTT device. In order to emulate an MQTT device, a user has to mention MQTT broker IP & port, username & password of the MQTT broker if exists and MQTT device type, i.e., subscriber or publisher.

If the device is a subscriber, then the user must mention the topic name which the device needs to be subscribed. For this purpose, a user can either select a topic from the topic list which the user has already saved or he/she may add a new topic if the topic name does not exist the topic list.

If the device is a publisher, then the user must mention the topic name along with the time profile and data profile in order to emulate the device behaviour.

- For the time profile, a user must tell in which frequency a device publishes the data, i.e., whether a device sends data after some specific interval (i.e., periodically, randomly, etc.) or after the occurrence of some event.

- For the data profile, a user must tell what type of data the device will is published by IoT device? It may be a numeric or binary or String data. Moreover, the user must define the data values or range of values that the device can publish in real-time in order to emulate the publishing device.

Both the time profile and data profile can be saved in the database so that they may be used later if some other device has the same time or data profile.

Likewise, if a user selects CoAP protocol from the main window, then the CoAP related GUI will be enabled and the user has to give information about CoAP server IP & port, CoAP method, time profile and data profile. The CoAP method includes four options, i.e., GET, POST, PUT and UPDATE which are briefly described in Table I.

*2) Business Logic Layer:* The business logic layer is responsible for implementing the user requirements that have been gathered from the presentation layer. We used four open-source APIs and database management system (DBMS) to implement the user required functionalities which include MQTT API [17], CoAP API [18] and libtins API [19]. The MQTT API [17] is used to create MQTT publisher and MQTT subscriber devices. An MQTT subscriber device will first send the subscribe request to the MQTT broker against some topic to establish the connection with MQTT broker and then will receive messages when some IoT device will publish a message to the MQTT broker against the subscribed topic. Similarly, an MQTT publisher device will first establish a connection with MQTT broker then will publish the message to the MQTT broker using MQTT API [17]. Likewise, the CoAP API [18] is used to send the user request to the CoAP server. The libtins API [19] is used to generate an attacking device. Finally, SQL DBMS is used to store the use case related information.

TABLE I
COAP METHODS

| Method | Description |
|---|---|
| GET | Retrieves the information corresponding to the URI request |
| POST | Requests the server to process the representation enclosed in the request |
| PUT | Requests to update or create a resource identified by URI |
| DELETE | Requests to delete the resource identified by the URI |

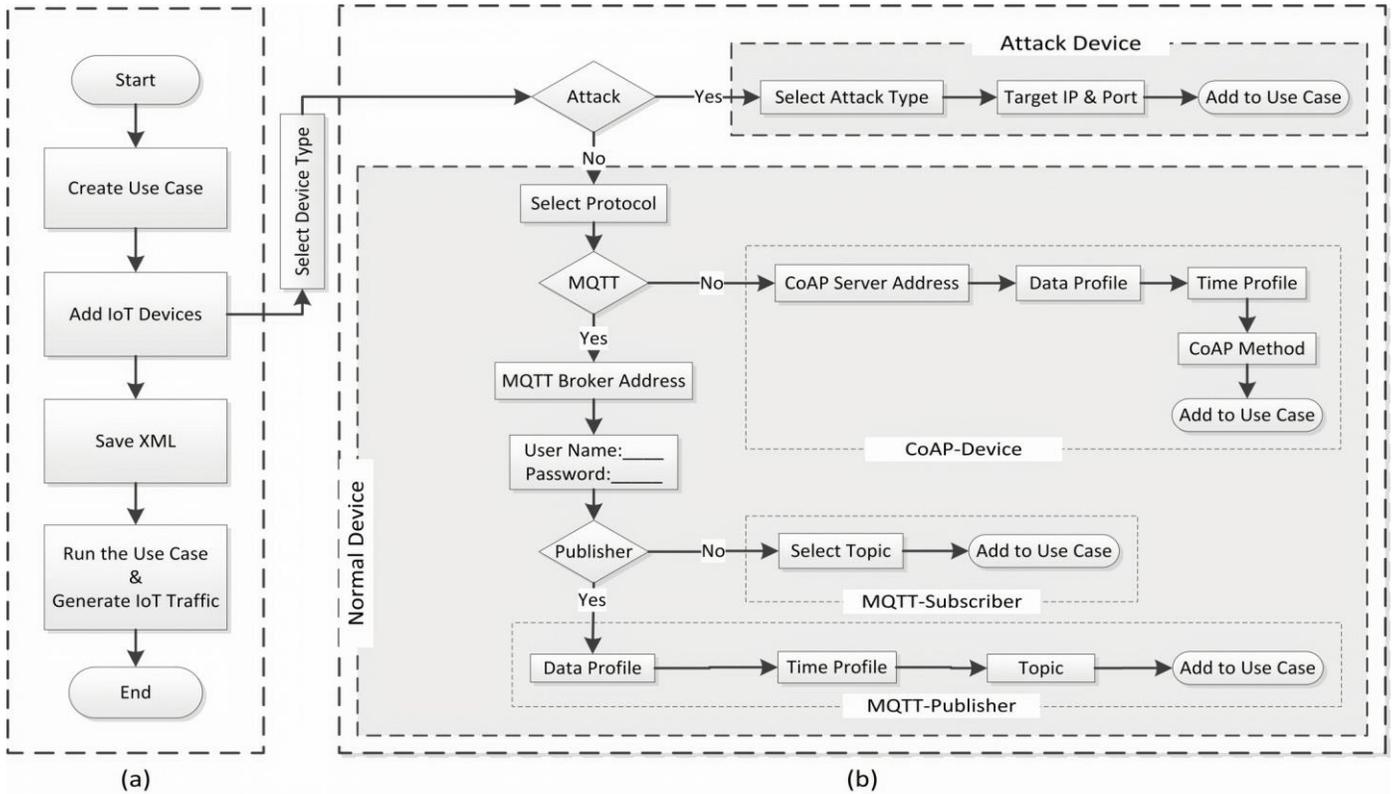

Fig. 2. Flow Diagram of IoT-Flock

*3) Network Layer:* The network interface layer is responsible for assigning a unique virtual IP to each IoT device from a single physical system. Moreover, it is also responsible for sending and receiving the data to/from the network.

### B. IoT-Flock Working

The flow diagram shown in Fig. 2(a) illustrates the working of IoT-Flock. It consists of four steps: create use case, add IoT devices to use case, save XML and run the use case & generate IoT traffic. Upon starting the IoT-Flock in GUI mode, a user will first create the desired use case by simply clicking at '+ new use case' button. After entering the use case name, the user will click on 'OK' to create the use case and the use case window will appear. Now the user has to add IoT devices into the use case. Fig. 2(b) shows the device creation steps. A user can create both IoT normal and attack device with respect to use case. After adding the devices, the next step is to generate an XML file of the use case. In case, if a user needs to re-run the use case after some time, then a user can simply load and run the saved XML file of the use case instead of re-creating it. Moreover, the generated XML file can be shared with anyone to ease him/her in understanding the use case. Lastly, the user will click on the 'Start' button to initiate the emulation of the use case.

### C. Attack Types in IoT-Flock

Currently, we included four recent IoT application layer vulnerability attacks in the IoT-Flock. However, one can add more types of attacks in IoT-Flock by extending the source code. These attacks are reported recently in a well-known vulnerability reporting platform, i.e., national vulnerability database (NVD) [20]. Below is the description of these attacks.

MQTT Packet Crafting Attack - In this attack, MQTT packets are specially crafted to crash an application. The attacker first establishes a connection with MQTT broker at Transport layer and then sends the MQTT publish command right at the beginning instead of sending a connection request to MQTT broker [21].

MQTT Publish Flood - IoT devices follow a periodic or event-driven model for sending data using application layer protocols. In the periodic model, the device is set to send data after every x interval, e.g., temperature sensor sends temperature data after every five seconds to the server and in event-driven model device send data when some event occurs e.g motion sensor is configured to only send data to the server when it detects motion in the environment. According to [22], MQTT publishing message at a high rate can cause a denial of service (DoS) attack.

TABLE II
ACTUAL TRAFFIC FEATURES

| Device Name | Average Value | Mode | Average Time | Behavior |
|---|---|---|---|---|
| Temp Sensor | 34.72 C | 35 C | 180s | Periodic |
| Light Sensor | 120 Lux | 100 Lux | 180s | Periodic |
| Motion Sensor | 0.5 | 1 | 3s-5s | Random |
| Humidity Sensor | 40 % | 42 % | 180s | Periodic |

TABLE III
EMULATED TRAFFIC FEATURES

| Device Name | Data Profile | Time Profile | Behavior |
|---|---|---|---|
| Temp Sensor | 33-35 C | 180s | Periodic |
| Light Sensor | 99-120 Lux | 180s | Periodic |
| Motion Sensor | 0-1 | 3s-5s | Random |
| Humidity Sensor | 39-42 % | 180s | Periodic |

CoAP Segmentation Fault Attack - While communicating with the CoAP server, a valid Uri-Path is an essential part of the request and response packet. Recently, an attack is reported in which the attacker when sets the Uri-Path as null, then CoAP server mishandles such packets hence causes the segmentation fault [23]. Moreover, an attacker can generate the DoS attack by sending such packets in a large amount.

CoAP Memory Leak Attack - CoAP server sends or receives data based upon the CoAP methods as called by the clients. It is reported that when an attacker sends invalid options to the CoAP server, it causes the memory crash as the processing of packet with single invalid option wastes 24 bytes of memory [24].

## IV. EXPERIMENTAL ANALYSIS

### A. Experimental Setup

*1) Use Case Creation & Traffic Generation:* A real-time IoT smart home system is deployed in our laboratory and office environment. In our smart home use case ten IoT devices are installed. Each device contains four types of environment monitoring sensors, i.e., temperature sensor, humidity sensor, motion sensor and light sensor. All the devices and sensors communicate over MQTT protocol within a wireless local area network (WLAN) environment. All the sensors except the motion sensors are set to send the data periodically to the server after every 3 minutes where this received data is saved into a database. The motion sensor sends data to the server only when it detects motion. The data saved in the server database is read by a rule engine which controls the switching of home appliances based on the conditions as defined by the user. In our case, the rule engine is controlling three types of appliances which includes air conditioners (AC), lights and fans.

To implement the above discussed real-time use case into IoT-Flock, we first analyzed the traffic patterns of the real-time devices as shown in Table II. For time profile, we observed that all sensors except the motion sensor are sending data periodically, i.e., after every 180s. So, we set their time profile as periodic at 180s and set data profile using (1):

$$Data\ Range = (MIN\ [Average\ Value, Mode\ Value],\\ MAX[Average\ Value, Mode\ Value]) \quad (1)$$

For motion sensor, we set it to send 1 and 0 randomly between time range 3s to 5s. Once we finalized the time profile and data profile of the smart home use case as shown in Table III, next we created a use case in IoT-Flock and named it as 'Smart Home' by following the steps shown in Fig. 2. Furthermore, we also created an attack network of 8 devices generating four types of attacks as mentioned in Section III-C by running IoT-Flock tool on a different machine. To discriminate the normal and attack traffic, we fixed two IPV4 address ranges, e.g., IP range from 192.168.1.2 to 192.168.1.254 fixed for normal network devices and IP range from 192.168.2.2 to 192.168.2.254 fixed for attacking network devices.

*2) Data Capturing:* After creating both the normal and attack networks, we first started the normal network to generate IoT normal traffic. After 10 minutes, we started the attack network. The whole emulation ran for 30 minutes and we captured both the normal and attack traffic using Wireshark [25] and saved it into a .pcap file for further analysis.

*3) Features Extraction:* Once we got the .pcap file of both normal and attack network, we then extracted the features from .pcap file to train the machine learning classifiers for attack detection. For this purpose, we used a publicly available tool, i.e., CICFlowmeter [26] to extract the network traffic flow features of the given .pcap file and saved it into a .csv file. The CICFlowmeter [26] extracts more than 80 features from a .pcap file. The further details of these features are mentioned in [27]. We extracted network traffic features from the .pcap file and labelled them with respect to normal and attack device IP addresses. The final .csv file had 400 samples of both normal and attack traffic.

*4) Features Selection:* Feature selection plays a significant role in the performance of a machine learning (ML) model as it selects the significant features which are imperative for data classification and ignores the useless features which can disturb the performance of ML model. For features selection, we used mutual information feature selection technique which calculates the mutual information among the features for a discrete target variable in order to select the important features from a dataset. We selected the top 10 features for better training and testing of the machine learning model.

*5) Model Training:* After selecting the features, we split the data into training and testing set by randomly splitting 20% data for testing and 80% for training. We then trained and tested three commonly used ML models, i.e., Naive Bayes (NB), Random Forest (RF) and K-Nearest Neighbor (KNN) over the pre-processed dataset.

### B. Performance Metrics

For the performance evaluation of the machine learning classifiers, we calculated three commonly used performance parameters over the testing data which include sensitivity, specificity and accuracy. The sensitivity is defined as the ability of the system to correctly detecting the attack. The specificity is defined as the ratio of normal packets that mistakenly are classified as malicious packet. The accuracy is defined as the ratio of correct predictions with respect to all samples. Mathematically, these are expressed in (2)-(4):

$$Sensitivity = \frac{TP}{TP + FN} \times 100 \quad (2)$$

$$Specificity = \frac{TN}{TN + FP} \times 100 \quad (3)$$

$$Accuracy = \frac{TP + TN}{TP + FN + TN + FP} \times 100 \quad (4)$$

TABLE IV
RESULTS

| Parameters | NB | KNN | RF |
|---|---|---|---|
| Sensitivity | 99.01 | 99.96 | 99.98 |
| Specificity | 95 | 99.88 | 99.99 |
| Accuracy | 97.14 | 99.92 | 99.99 |

Table IV summarizes the overall results of the three machine learning classifiers which are used for detecting malicious traffic.

## V. CONCLUSION

Security is the major concern which may cramp the proliferation of IoT devices. Network traffic generation is a key technique which is used to design and analyze the performance of network security solutions like firewall, intrusion detection system (IDS), intrusion prevention system (IPS), etc. In this work, we proposed an open-source traffic generation framework, i.e., IoT-Flock which supports two IoT application layer protocols. Most of the existing open-source traffic generators lack of generating the attack traffic, however, IoT-Flock can generate both the attack and normal IoT traffic in order to train and test the IoT network security solutions. To demonstrate how the IoT-Flock can help in generating an IoT security solution, we first analyzed the traffic behaviour of a real-time smart home use case. We then created the smart home use case in IoT-Flock. After that, we created an attacking network to attack the IoT devices using IoT-Flock and captured the traces of both normal and attack traffic. Finally, we extracted the features then trained and tested the three commonly used machine learning algorithms for detecting the malicious traffic in IoT smart home use case. Among these models, Random Forest classifier performed the best with an accuracy of 99.99%. One can download IoT-Flock from GitHub [28], create real-time IoT use case, generate the use case traffic and test or design security solutions for IoT use case by utilizing IoT-Flock. Currently, IoT-Flock supports four recently reported IoT attacks, however, one can add more IoT attacks in the tool by extending the source code of the tool.